\documentclass[floats,floatfix,showpacs,amssymb,prd,superscriptaddress,twocolumn,aps,nofootinbib]{revtex4}
\usepackage{graphicx, epsfig, amssymb} 
\usepackage{amsmath, amsfonts}
\usepackage{bm} 

\usepackage[linktocpage]{hyperref}
\usepackage[caption=false]{subfig}
\usepackage[usenames]{color}

\newcommand{\ee}{\end{equation}}
\newcommand{\eea}{\end{eqnarray}}
\newcommand{\be}{\begin{equation}}
\newcommand{\bea}{\begin{eqnarray}}

\begin{document}

\title{Kerr black holes with self-interacting scalar hair: \\ hairier but not heavier}

 \author{Carlos~A.~R.~Herdeiro, Eugen~Radu and Helgi\ R\'unarsson}
   \affiliation{
   Departamento de F\'\i sica da Universidade de Aveiro and CIDMA, 
   Campus de Santiago, 3810-183 Aveiro, Portugal.
 }


\date{September 2015}

\begin{abstract}
The maximal ADM mass for (mini-)boson stars (BSs) -- gravitating solitons of Einstein's gravity minimally coupled to a free, complex, mass $\mu$, Klein-Gordon field -- is $M_{\rm ADM}^{\rm max}\sim M_{Pl}^2/\mu$. Adding quartic self-interactions to the scalar field theory, described by the Lagrangian $\mathcal{L}_I=\lambda |\Psi|^4$, the maximal ADM mass becomes $M_{\rm ADM}^{\rm max}\sim \sqrt{\lambda}M_{Pl}^3/\mu^2$. Thus, for mini-BSs, astrophysically interesting masses require ultra-light scalar fields, whereas self-interacting BSs can reach such values for bosonic particles with Standard Model range masses. We investigate how these same self-interactions affect Kerr black holes with scalar hair (KBHsSH)~\cite{Herdeiro:2014goa}, which can be regarded as (spinning) BSs in stationary equilibrium with a central horizon. Remarkably, whereas the ADM mass scales in the same way as for BSs, the \textit{horizon mass} $M_H$  does not increases with the coupling $\lambda$, and, for fixed $\mu$, it is maximized at the ``Hod point", corresponding to the extremal Kerr black hole obtained in the vanishing hair limit. This mass is always $M_{\rm H}^{\rm max}\sim M_{\rm Pl}^2/\mu$. Thus, introducing these self-interactions, the black hole spacetimes may become considerably ``hairier" but the trapped regions cannot become ``heavier". We present evidence this observation also holds in a model with $\mathcal{L}_I= \beta|\Psi|^6-\lambda|\Psi|^4$; if it extends to \textit{general} scalar field models, KBHsSH with astrophysically interesting horizon masses \textit{require} ultra-light scalar fields. Their existence, therefore, would be a smoking gun for such (beyond the Standard Model) particles. 

\end{abstract}




\maketitle

 
 \newpage

\section{Introduction}

Boson stars (BSs) -- see~\cite{Schunck:2003kk,Liebling:2012fv} for reviews --  were originally discovered in Einstein's gravity minimally coupled to a massive (mass $\mu$), free, complex scalar field~\cite{Kaup:1968zz,Ruffini:1969qy}. They are often described as macroscopic quantum states, or Bose-Einstein condensates, which are prevented from collapsing by Heisenberg's uncertainty principle. But a purely macroscopic interpretation can be obtained by regarding the scalar field as a perfect fluid, which, generically, has pressure (see $e.g.$~\cite{Faraoni:2012hn}). Naturally, the pressure can only prevent collapse into a black hole (BH) up to some maximal mass, as it is also the case for fermionic compact stars, where one obtains the Tolman-Oppenheimer-Volkoff limit. From generic arguments, the maximum ADM mass of BSs supported by a free complex scalar field is of the order of the Compton wavelength of the scalar field:
\begin{equation}
 M_{\rm ADM}^{\rm max}\simeq \alpha_{\rm BS} \frac{M_{\rm Pl}^2}{\mu}\simeq \alpha_{\rm BS} \, 10^{-19}M_\odot\left(\frac{\rm GeV}{\mu}\right)\ , 
\label{mini}
\end{equation}
 where $M_{\rm Pl},M_{\odot}$ are the Planck and Sun masses, respectively. The constant $\alpha_{\rm BS}$ must be obtained from computing the explicit solutions (numerically, as there are no closed form known solutions in four spacetime dimensions), and is of order unity. Its specific value depends on the `quantum' numbers of the BS. BSs can have a number of nodes $n\in \mathbb{N}_0$ in the scalar field profile and an azimuthal harmonic dependence $\Psi\sim e^{im\phi}$, with $m\in \mathbb{Z}^+$, if they are rotating. For instance, for nodeless BSs -- usually regarded as the fundamental and more stable states -- with azimuthal harmonic index $m=0$ (spherically symmetric), $\alpha_{\rm BS}=0.633$~\cite{Liebling:2012fv}, whereas for rotating solutions with $m=1;2$, $\alpha_{\rm BS}=1.315;2.216$~\cite{Yoshida:1997qf,Grandclement:2014msa}. As a rule of thumb, one may expect $M^{\rm max}_{\rm ADM}$  to increase with $m$ as $\sim m+1$. In any case, unless one considers absurdly high values of $m$, for typical standard model particle masses, say $\mu\sim 1$ GeV, this maximal mass is small, by astrophysical standards: $M_{\rm ADM}^{\rm max}\sim 10^{-19} M_\odot$. For this reason such BSs have been dubbed \textit{mini-}BSs. 
 
It was, however, shown by Colpi, Shapiro and Wasserman~\cite{Colpi:1986ye}, that $M^{\rm max}_{\rm ADM}$ can be considerably increased if quartic-self-interactions are introduced as a potential term: $V(|\Psi|)=\lambda|\Psi|^4$, where $\lambda$ is a coupling constant. Then, for spherically symmetric \textit{quartic}-BSs~\cite{Colpi:1986ye}
\begin{equation}
 M_{\rm ADM}^{\rm max}\simeq 
0.062  \sqrt{\lambda}\frac{M_{\rm Pl}^3}{\mu^2}\simeq 0.062  \sqrt{\lambda}M_\odot \left(\frac{\rm GeV}{\mu}\right)^2\ .
\label{csw}
\end{equation}
Thus, if, say, $\mu\sim 1$ GeV and $\lambda\sim 1$, this maximal mass is comparable to that of astrophysical objects: $M_{\rm ADM}^{\rm max}\sim 0.062 M_\odot$ and it grows without bound as $\lambda$ increases. In fact, as observed in~\cite{Colpi:1986ye}, $M^{\rm max}_{\rm ADM}$ becomes comparable to the Chandrasekhar mass for fermions of mass $\sim\lambda^{-1/4}\mu$.

Relation (\ref{csw}) was established by putting together numerical data and an analysis of the large $\lambda M_{\rm Pl}^2/\mu^2$ limit, where the equations simplify and some analytical relations can be obtained. In particular it was observed that when this quantity is large, the spatial
distribution of the quartic-BSs becomes more extended than for mini-BSs, with a slower fall off of the scalar field close to the centre of the star matching the asymptotic exponential fall off. Thus, quartic-BSs become larger than mini-BSs and can withstand larger masses without gravitationally collapsing into BHs.

A small number of works generalized quartic-BSs to the rotating case. The pioneering work by Ryan~\cite{Ryan:1996nk} focused on the aforementioned large self-interaction limit, whereas more recent works~\cite{Grandclement:2014msa,Kleihaus:2015iea} focused on a single value of the coupling $\lambda$ and obtained, for instance, solutions with higher $m$. A first goal of this work is to present a more
 systematic study of  rotating quartic-BSs in terms of $\lambda$. In particular, we obtain the analogue of (\ref{csw}) for solutions with $m=1$, which is ($cf.$ Fig.~\ref{max_mass_0}):
\begin{equation}
M_{\rm ADM}^{\rm max}\simeq 
0.057  \sqrt{\lambda}\frac{M_{\rm Pl}^3}{\mu^2}\simeq 0.057  \sqrt{\lambda}M_\odot \left(\frac{\rm GeV}{\mu}\right)^2\ .
\label{quartic_rotating}
\end{equation}
 
Our main interest on this maximal mass has, however, a different focus. In a recent development~\cite{Herdeiro:2014goa,Herdeiro:2015gia}, it was shown that one can add a BH at the centre of asymptotically flat mini-BSs, as a stationary solution, as long as the BH and the BS are in a synchronized co-rotation (see also~\cite{Dias:2011at}). These solutions form a family of \textit{Kerr black holes with scalar hair} (KBHsSH), since they interpolate continuously between mini-BSs -- in the limit of vanishing horizon -- and Kerr BHs -- in the limit of vanishing scalar field. The necessity for rotation is clearly understood in the Kerr limit, wherein the backreacting scalar field ``hair" reduces to test field ``clouds" around Kerr BHs, first described by Hod for extremal Kerr backgrounds~\cite{Hod:2012px} (see also~\cite{Hod:2013zza,Herdeiro:2014goa,Hod:2014baa,Benone:2014ssa,Hod:2014npa,Hod:2015ota}). These clouds are bound states at the threshold of the superradiant instability of Kerr BHs triggered by a massive bosonic field, which only exists for the Kerr family if the angular momentum is non-vanishing. Thus, it was suggested in~\cite{Herdeiro:2014goa,Herdeiro:2014ima} that any field triggering a superradiant instability (see~\cite{Brito:2015oca} for a review) of a (bald) BH background should allow the construction of hairy solutions with that field.\footnote{Some partial evidence for this conjecture was recently presented~\cite{Ponglertsakul:2015bpa}, in the context of charged BHs. For asymptotically flat Reissner-Nordstr\"om BHs there are no superradiant instabilities~\cite{Hod:2013eea,Hod:2013nn,Degollado:2013eqa}, unless the BHs are enclosed in a cavity~\cite{Herdeiro:2013pia,Hod:2013fvl,Degollado:2013bha} (see also~\cite{Li:2014xxa,Li:2014fna,Li:2015mqa}). In~\cite{Ponglertsakul:2015bpa} the corresponding Reissner-Nordstr\"om BHs with scalar hair enclosed in a cavity were constructed and evidence for their stability was presented.} 
In fact, an extra condition is that the field must give rise to a time-independent energy-momentum tensor from the threshold modes~\cite{Herdeiro:2015gia}. This explains why a single real scalar field cannot originate stationary scalar hair, a fact recently proved with generality~\cite{Graham:2014ina}. The full domain of existence of KBHsSH was exhibited in~\cite{Herdeiro:2014goa}; in particular it was shown that the ADM mass of these BHs is bound by the mass of the corresponding mini-BSs that are obtained in the limit of vanishing horizon. Hence, in this model, the maximum ADM mass of KBHsSH  is still given by (\ref{mini}), with the value of $\alpha_{\rm BS}$ for the appropriate $m\neq 0$.

One may ask, instead, what is the maximal \textit{horizon}, rather than ADM, mass for KBHsSH. The former is well defined in terms of a Komar integral. A first analysis of horizon mass and angular momentum for KBHsSH was presented in~\cite{Herdeiro:2015moa} where it was pointed out that the Kerr  bound can be violated in terms of horizon quantities. The second goal of this paper is to make a different observation: the maximum horizon mass $M^{\rm max}_H$ for KBHsSH is attained at a particular configuration we dub the \textit{Hod point}, corresponding to the extremal (vacuum) Kerr BH obtained in the vanishing hair limit, and precisely one of the points studied in~\cite{Hod:2012px}. For KBHsSH this maximal mass obeys the same form as \eqref{mini},
\begin{equation}
 M_{\rm H}^{\rm max}\simeq \alpha_{\rm Hod} \frac{M_{\rm Pl}^2}{\mu}\simeq \alpha_{\rm Hod} \, 10^{-19}M_\odot\left(\frac{\rm GeV}{\mu}\right)\ , 
\label{hodpoint}
\end{equation}
but with the constant $\alpha_{\rm Hod}<\alpha_{\rm BS}$. For instance, for $m=1;2$, $\alpha_{\rm Hod}=0.526; 1.165$.

Since the introduction of self-interaction changes the maximal ADM mass of BSs from  \eqref{mini} into \eqref{csw} or \eqref{quartic_rotating}, a natural question is what happens both to the ADM and horizon maximal mass for the case of the hairy BHs in the self-interacting theory.  For simplicity we shall henceforth dub these solutions as \textit{quartic-KBHsSH}, which have been recently considered in~\cite{Kleihaus:2015iea},  in a more general context.  
These authors presented the full domain of existence of quartic-KBHsSH, for values of $m$ up to 8, albeit for a single value of the coupling $\lambda$. Their domains of existence follow a similar pattern to the ones of KBHsSH and in particular their ADM mass is bounded by that of the corresponding BSs; this suggests formula~\eqref{quartic_rotating} applies also to quartic-KBHsSH. As a third goal of this paper, we generalize this observation for more generic couplings: we shall compute the domain of existence of quartic-KBHsSH for various values of the coupling $\lambda$ and, in particular, establish that (\ref{quartic_rotating})  still provides the maximum ADM mass of quartic-KBHsSH for coupling $\lambda$ (and $m=1$, $n=0$). 
 
 A quite different behaviour is observed for the maximal \textit{horizon} mass of quartic-KBHsSH. The fourth goal of this paper is to establish that, regardless of the coupling in the self-interacting case, the maximal horizon mass is still attained at the Hod point. Thus, relation (\ref{hodpoint}) provides the maximal horizon mass of quartic-KBHsSH for coupling $\lambda$. Consequenty, as $\lambda$ increases, the spacetime of quartic-KBHsSH can become considerably hairier, in the sense that the ratio between $M_{\rm ADM}^{\rm max}/M_{\rm H}^{\rm max}$ can increase dramatically; but the BH itself, $i.e.$ the horizon, cannot become ``heavier", $i.e.$ more massive.
 
 This paper is organized as follows. In Sec.~\ref{sec_model} we briefly describe the quartic model. In Sec.~\ref{sec_II}  we present the quartic-BS solutions, emphasizing the variation of the maximal mass with the coupling. In Sec.~\ref{sec_III} we describe the quartic-KBHsSH in terms of ADM quantities, giving the full domain of existence for three different values of the self-coupliing $\lambda\neq 0$, which we compare with the non-self-interacting case ($\lambda=0$).  In Sec.~\ref{sec_IV} we explore the horizon quantities for the original and quartic-KBHsSH, showing that the maximum horizon mass is reached at the Hod point. Finally, in Sec.~\ref{sec_V} we discuss the universality of our main observation, which we support by examining \textit{Q}-KBHsSH (defined therein)  and discuss some possible research directions related to phenomenology.
 
 \newpage

\section{The model}
\label{sec_model}
 
We shall consider a massive, complex scalar field, $\Psi$, minimally coupled to Einstein's gravity. The action is:
\begin{eqnarray}
  \label{action}
  &&
	\mathcal{S} = \int d^4x \sqrt{-g}\left[\frac{R}{16\pi G}- g^{ab}\Psi^*_{,a}\Psi_{,b} - U(|\Psi|)\right],~~{~~} \ \  
\\
\label{pot}
&&~~~~{\rm with}~~~U(|\Psi|)= \mu^2\left|\Psi\right|^2 + \lambda\left|\Psi\right|^4,
\end{eqnarray}  
where $G$ is Newton's constant, $\mu$ is the scalar field mass and the self-coupling is positive, $ \lambda>0$. 
To describe both spinning BSs and KBHsSH we take the metric ansatz \cite{Herdeiro:2014goa}
%
\begin{eqnarray}
  ds^2 = -e^{F_0}Ndt^2 + e^{2F_1}\left( dr^2/N + r^2d\theta^2 \right) \nonumber \\ + e^{2F_2}r^2\sin^2\theta \left(               d\varphi - Wdt \right)^2 \ ,
\end{eqnarray}
where $N\equiv 1-r_H/r$, $F_i,W$, $i=1,2,3$ are functions of the spheroidal coordinates $r,\theta$. The scalar field ansatz is
\begin{eqnarray}
\label{scalar}
  \Psi = e^{-iwt+im\varphi}\phi(r,\theta) \ .
\end{eqnarray}
$r_H$ is the radial coordinate of the horizon in the BH case and is set to zero for BSs. $w$ is the scalar field frequency, $m\in \mathbb{Z^+}$ is the azimuthal harmonic index that will always be taken $m=1$ in the work. Also, we consider only nodeless solutions ($n=0$), meaning that the scalar field profile $\phi(r,\theta=\pi/2)$ has no zeros.

The Einstein-Klein-Gordon equations obtained from \eqref{action} yield a system of five coupled, non-linear PDEs, together with two constraint equations, found 
by taking $\mu^2\to \mu^2+ \lambda \phi^2$ ($\mu^2\to \mu^2+2\lambda \phi^2$) in the Einstein (Klein-Gordon) equations, as displayed in~\cite{Herdeiro:2015gia}. 
The boundary conditions used to solve these
equations are different depending on considering BSs or KBHsSH, but do not depend on $\lambda$, and can also be found in~\cite{Herdeiro:2015gia}, together with a description of the numerical code used to solve the PDEs, employing a Newton-Raphson relaxation method.

\section{Quartic-Boson Stars}
\label{sec_II}
\begin{figure}[h!]
  \begin{center}
    \includegraphics[width=8.9cm]{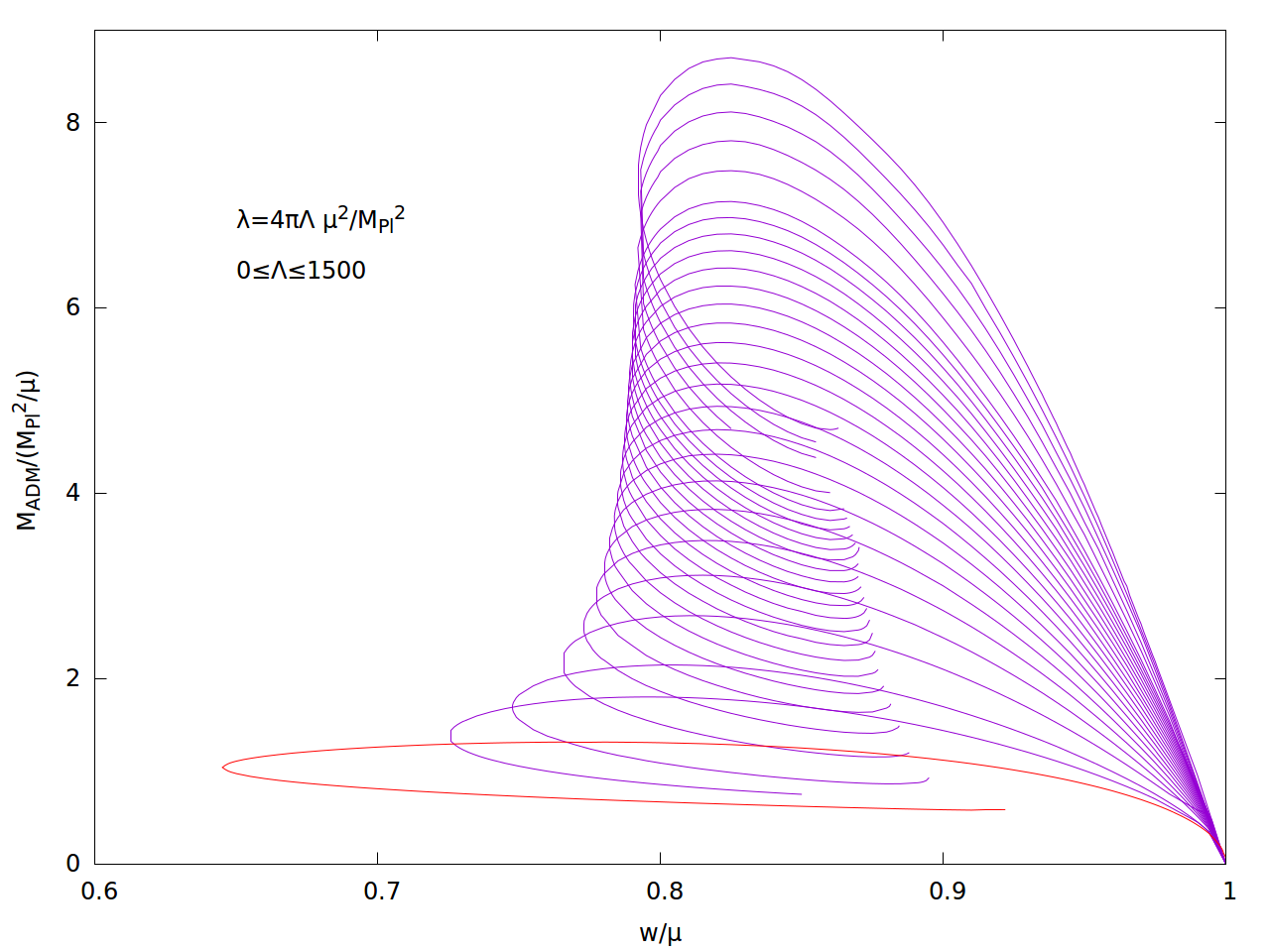}
  \end{center}
  \caption{ADM mass of $m=1$, nodeless BSs, as a function of $w$, for various values of $\lambda$ (defined by $\Lambda$, used in numerics): (from bottom to top) $\Lambda=0$ (red curve), $\Lambda=25$, $\Lambda=50-1000$ in steps of $50$, and $\Lambda=1000-1500$ in steps of $100$. Each curve was drawn from, typically, one hundred solution points.}
  \label{fig:w-vs-M-BSs}
\end{figure}

We start by considering quartic-BS solutions and their dependence on the coupling $\lambda$. These solutions are  conveniently presented in an ADM mass $vs.$ scalar field frequency diagram, as shown in Fig. \ref{fig:w-vs-M-BSs}. In this type of diagram, for $\lambda=0$, it is well known that the BS solutions form a spiraling curve (see $e.g.$~\cite{Herdeiro:2015gia}). The solutions start from vacuum ($w=\mu$), at which point they trivialize. As $w$ decreases from this value, their ADM mass increases until reaching a maximum value at $w_{\rm max}\simeq 0.775 \mu$. A minimum frequency is then reached after which the curve backbends towards a second branch. A second backbending and a third branch can be seen in  Fig. 1 of~\cite{Herdeiro:2015gia}.

A similar description applies for $\lambda\neq 0$. In Fig. \ref{fig:w-vs-M-BSs} we can see how BSs change as the quartic coupling parameter $\lambda$ varies from zero -- the mini-BS case -- to a large value. The typical trend is that the spiral-like behaviour is still kept as $\lambda$ increases, with an increase of the maximal mass. Also, the range of frequencies becomes slightly smaller with increasing coupling. Each curve in Fig. \ref{fig:w-vs-M-BSs} continues after the local minimum of the mass towards a third branch which has been omitted for simplicity, but can be observed in the plots of the next section. Finally, observe that all BS curves merge at the origin, where they trivialize.

\begin{figure}[h!]
  \begin{center}
    \includegraphics[width=9cm]{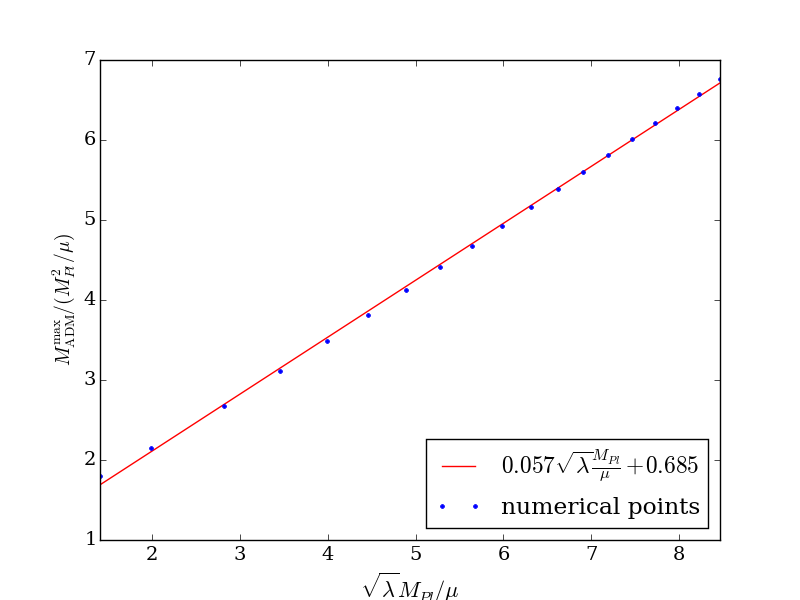}
  \end{center}
  \caption{$M_{\rm ADM}^{\rm max}$ for BSs $vs.$ the coupling constant $\lambda$, using the data in Fig.~\ref{fig:w-vs-M-BSs}, together with a numerical fit.}
  \label{max_mass_0}
\end{figure}

From the data exhibited in Fig.~\ref{fig:w-vs-M-BSs} one obtains the  maximal mass $M_{\rm ADM}^{\rm max}$ $vs.$ $\lambda$ exhibited in Fig.~\ref{max_mass_0}; for large $\lambda$ the data points are well fitted by the formula announced in the Introduction: \eqref{quartic_rotating}.
%

\section{Quartic-KBHsSH: ADM quantities}
\label{sec_III}
\begin{figure}[h!]
  \begin{center}
     \includegraphics[width=8.55cm]{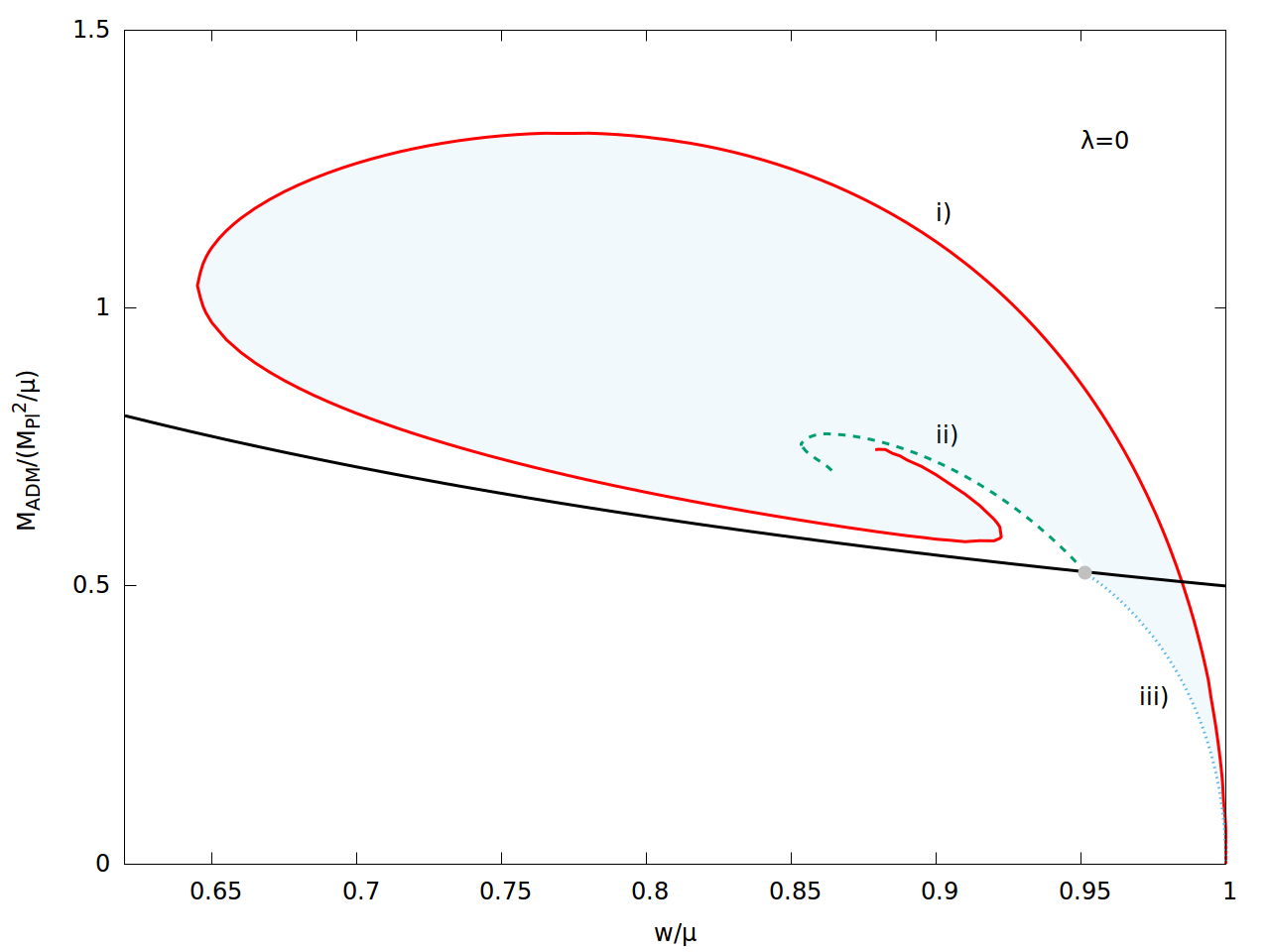}
      \end{center}
  \caption{Domain of existence of KBHsSH without self-interactions (blue shaded region).}
  \label{kbhsh0}
\end{figure}

We now turn to KBHsSH and consider first no self-interactions. For nodeless, $m=1$ KBHsSH, their domain of existence, in an ADM mass $vs.$ $w$~\cite{Herdeiro:2014goa,Herdeiro:2015gia} is shown in Fig.~\ref{kbhsh0}.  The black solid line represents extremal BHs {with horizon angular velocity}
\begin{eqnarray}
\label{cond}
 \Omega_H=\frac{w}{m} \ .
\end{eqnarray} 
We recall this is the synchronization condition which underlies KBHsSH. (Vacuum) Kerr BHs exist in this diagram below this line. The domain of existence of KBHsSH is bounded by:
\begin{description}
\item[i)] the corresponding BS curve (red solid line), for which we display the first three branches;
\item[ii)] by a line of extremal ($i.e.$ zero temperature) KBHsSH (green dashed line);
\item[iii)] and by a line of (vacuum) Kerr BHs (blue dotted line), corresponding to the Kerr solutions that allow the existence of stationary scalar clouds with the appropriate set of quantum numbers~\cite{Herdeiro:2014goa,Benone:2014ssa}.   
\end{description}
The ``Hod point" (grey dot) lies at the intersection of boundaries ${\bf ii)}$ and ${\bf iii)}$, corresponding to the extremal (vacuum) Kerr BH obtained in the limit of vanishing scalar field. Remarkably,  at the Hod point, the scalar field possesses a relatively simple closed form~\cite{Hod:2012px}.

\begin{figure}[h!]
  \begin{center}
    \includegraphics[width=8.55cm]{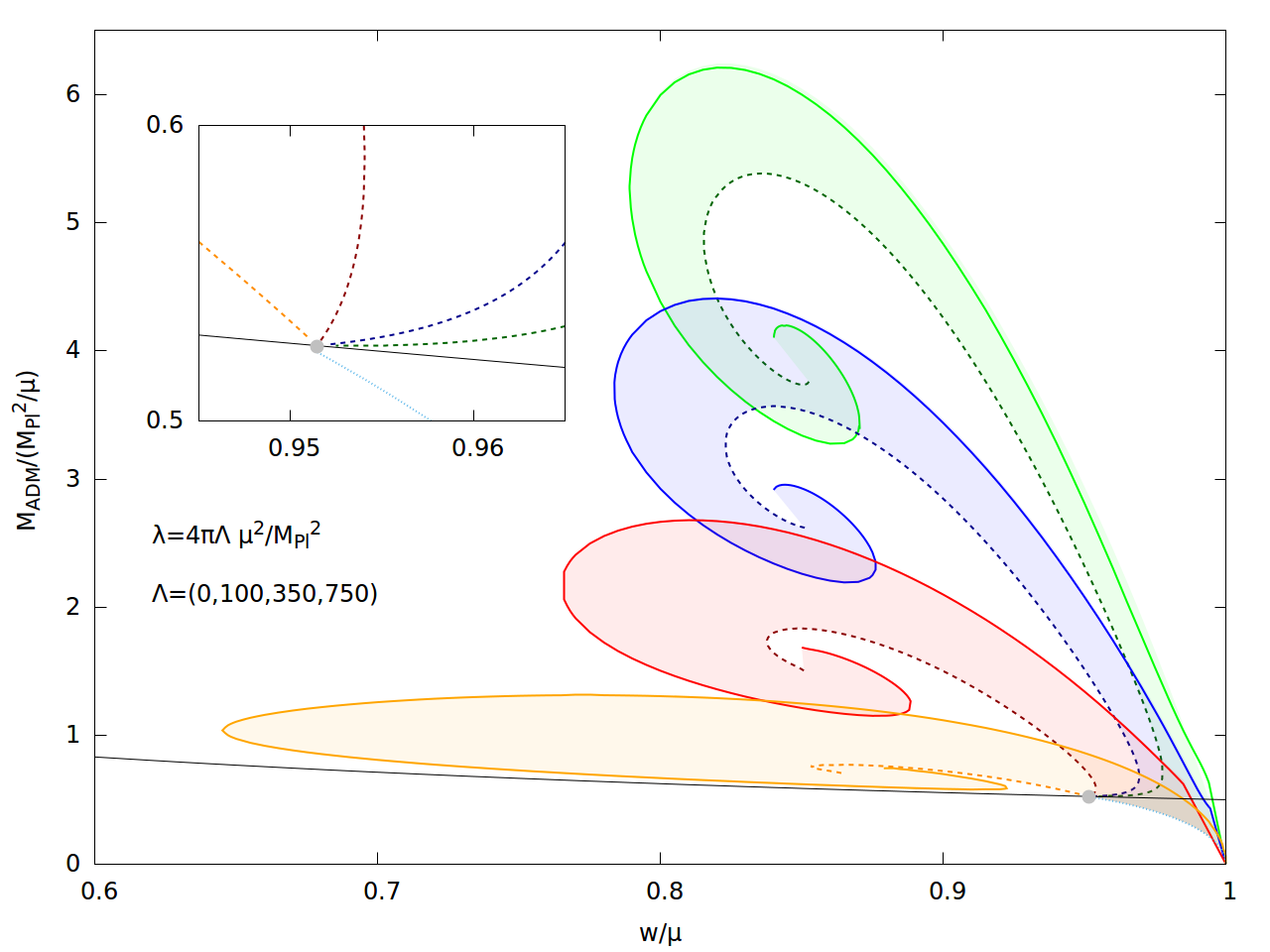}
      \end{center}
  \caption{Domain of existence of KBHsSH for $\Lambda=0,100,350,750$ (shaded regions, from the lowest to the highest, respectively). The inset shows a detail around the Hod point (grey dot).}
  \label{kbhsh1}
\end{figure}

When a quartic self-interaction is included for the scalar field, it was already shown in~\cite{Kleihaus:2015iea} that a similar pattern occurs, for a particular value of the coupling. In Fig.~\ref{kbhsh1} we show the domain of existence of KBHsSH for four different values of the coupling $\lambda$, given by $\Lambda=0,100,350,750$. Each domain of existence has been filled in by thousands of numerical points. As can be observed, as the coupling is varied, boundaries  ${\bf i)}$ and ${\bf ii)}$ change, and seem to be delimiting a gradually thinner ribbon in this diagram. Boundary  ${\bf iii)}$, however, is invariant, and so is the Hod point, as the self-interaction of the scalar field become irrelevant in the probe limit. In particular, this analysis confirms that the maximal ADM mass for quartic-KBHsSH is that of the corresponding BSs, and so it is still given by~\eqref{quartic_rotating}.

\begin{widetext}

\begin{figure}[h!]
  \begin{center}
    \includegraphics[width=8.55cm]{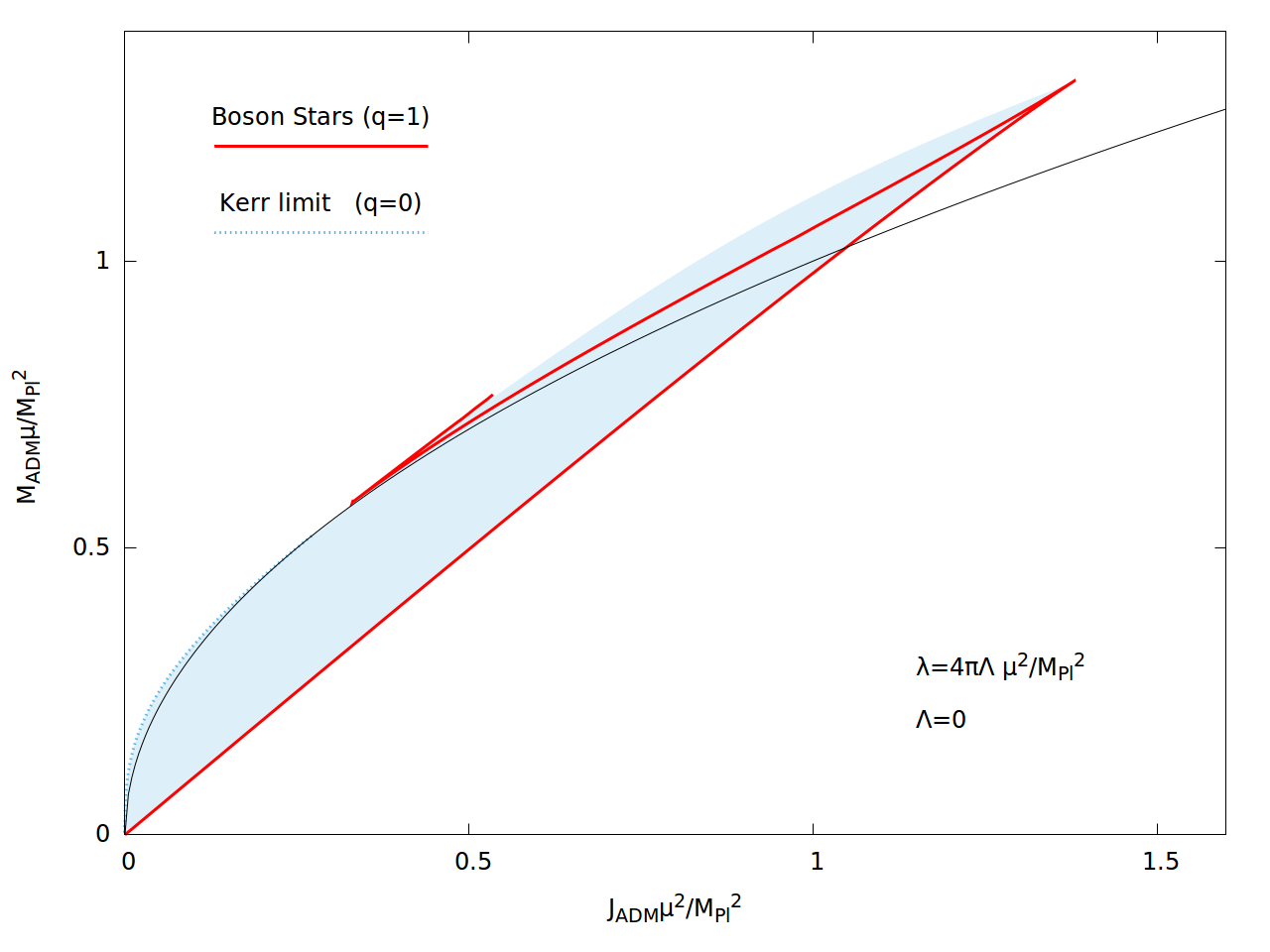}
     \includegraphics[width=8.55cm]{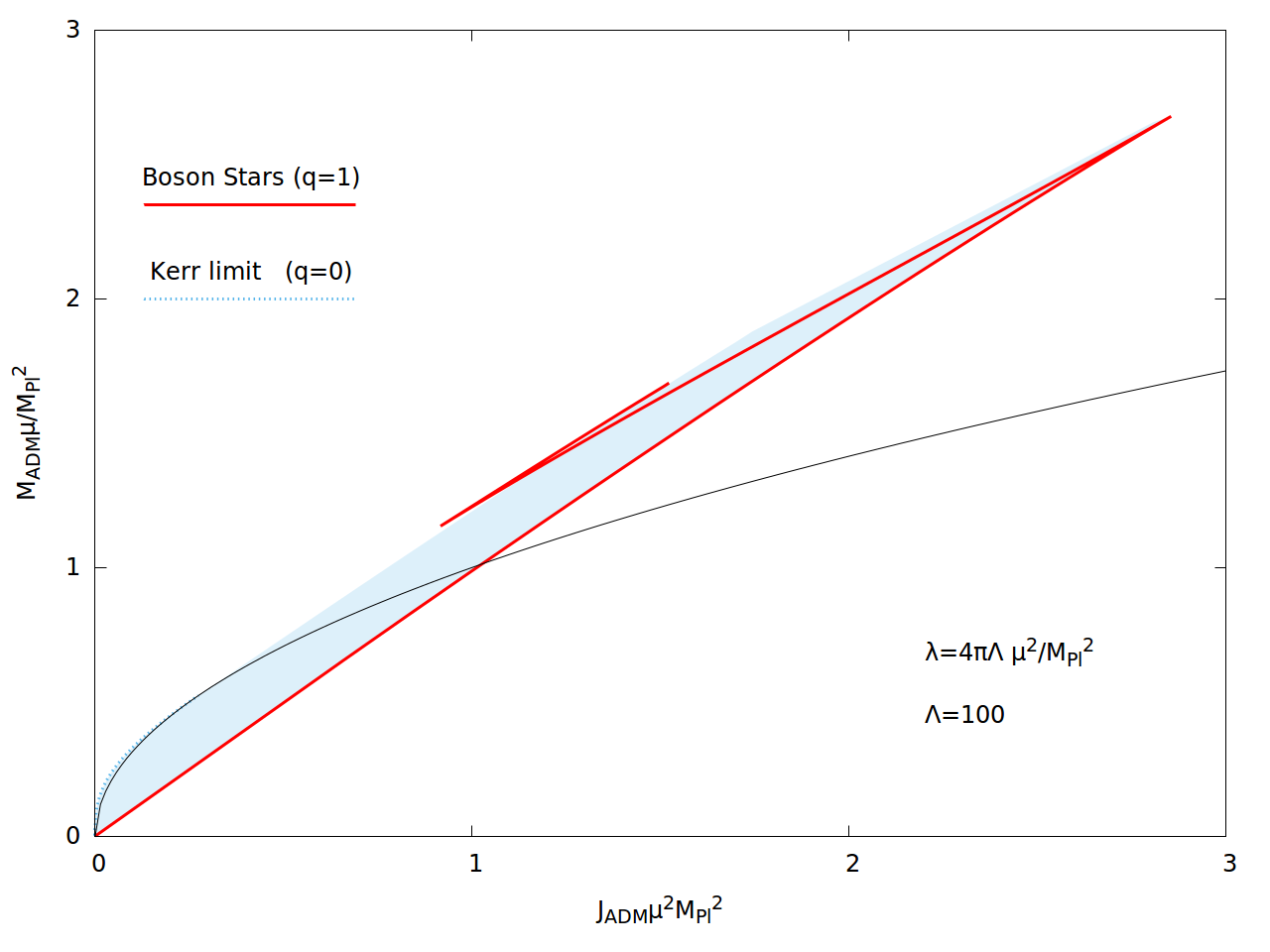}
      \includegraphics[width=8.55cm]{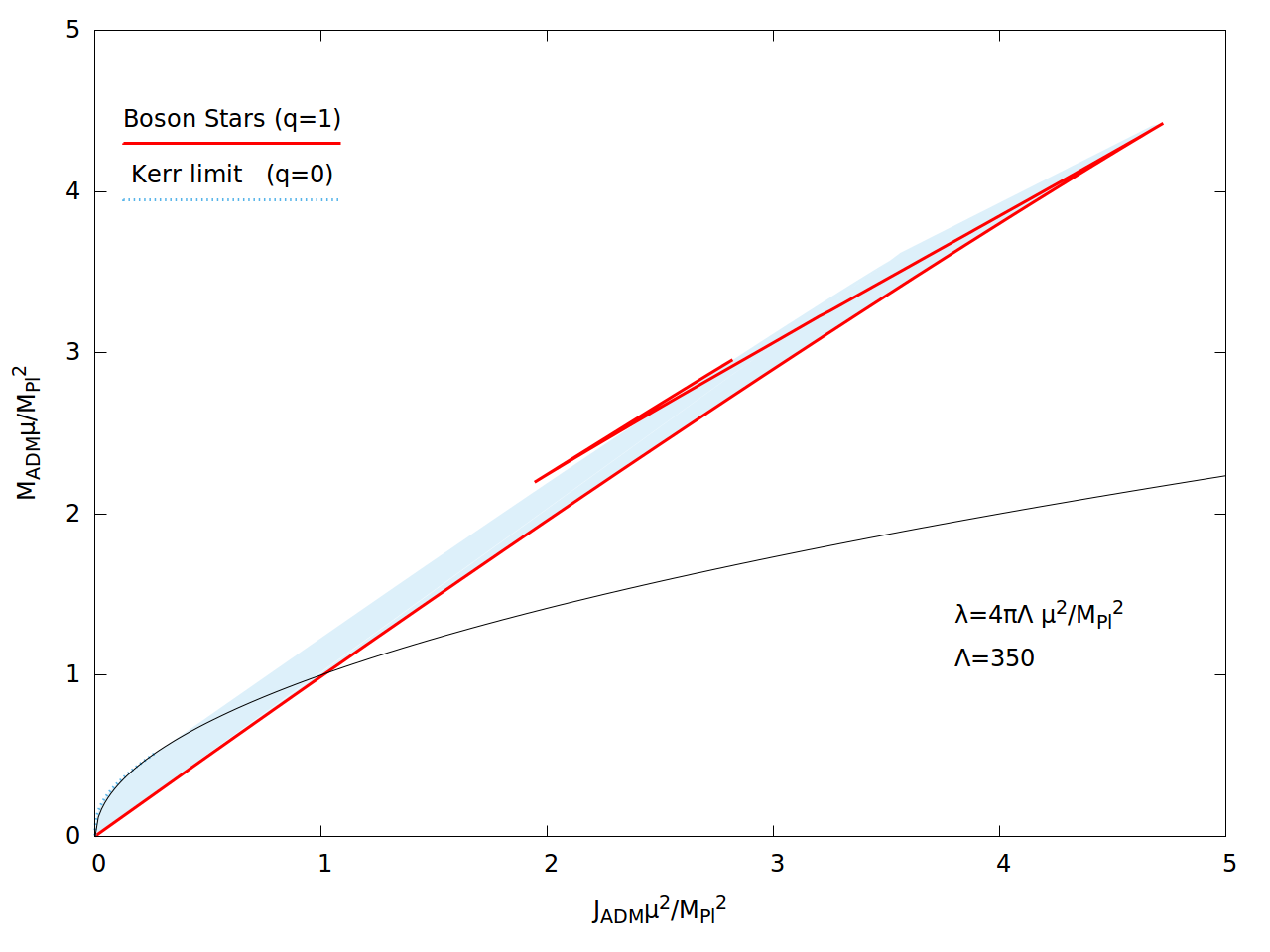}
       \includegraphics[width=8.55cm]{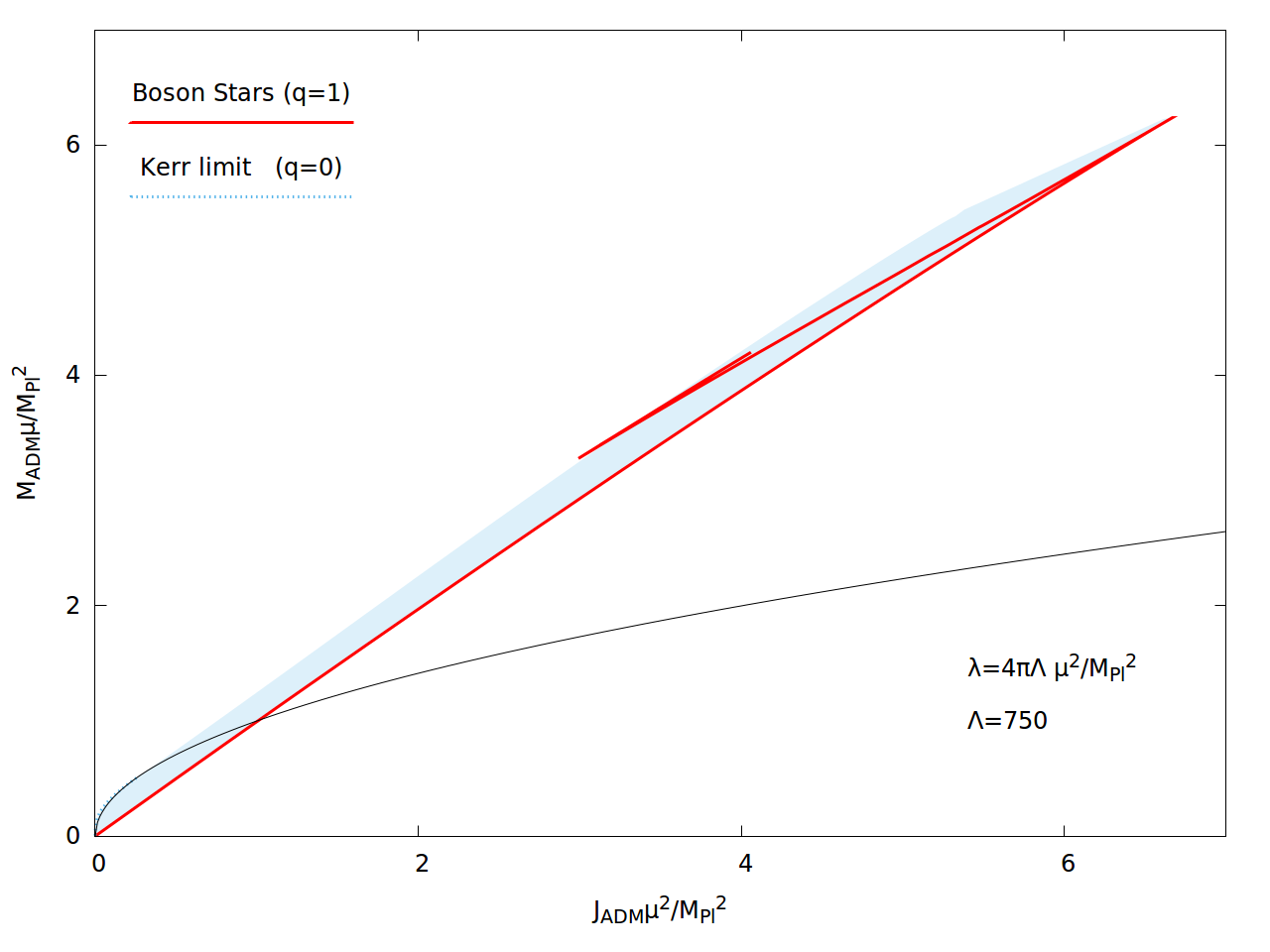}
  \end{center}
  \caption{Domain of existence of KBHsSH in an ADM mass $vs.$ ADM angular momentum diagram, for $\Lambda=0,100$ (top left and right panels) and $\Lambda=350,750$ (bottom left and right panels).}
  \label{fig:no-HBHs}
\end{figure}

\end{widetext}

In Fig.~\ref{fig:no-HBHs} we exhibit the phase space in terms of ADM quantities for quartic-KBHsSH for the four values of the coupling also used above. As before, the red solid line denotes BSs, the black solid line marks extremal Kerr (which means that Kerr BHs, exist \textit{above} this line, in this diagram), the blue dotted line is the zero mode line and blue shaded region is where (quartic-)KBHsSH exist. Analogously to the $\lambda=0$ case, there is always a positive correlation between the ADM mass and the ADM angular momentum, $J_{ADM}$, for the self-interacting case. In fact, the diagram looks quite similar, regardless of $\lambda$, with the major difference that, as $\lambda$ increases, larger values of $M_{\rm ADM}$ and $J_{ADM}$ are allowed. Also observe that for all cases there are solutions violating the Kerr bound, in terms of ADM quantities, as noticed for KBHsSH in~\cite{Herdeiro:2014goa,Herdeiro:2015moa} and for a specific example of quartic-KBHsSH in~\cite{Kleihaus:2015iea}.

\section{Quartic-KBHsSH: horizon quantities}
\label{sec_IV}

\begin{figure}[h!]
  \begin{center}
    \includegraphics[width=8.55cm]{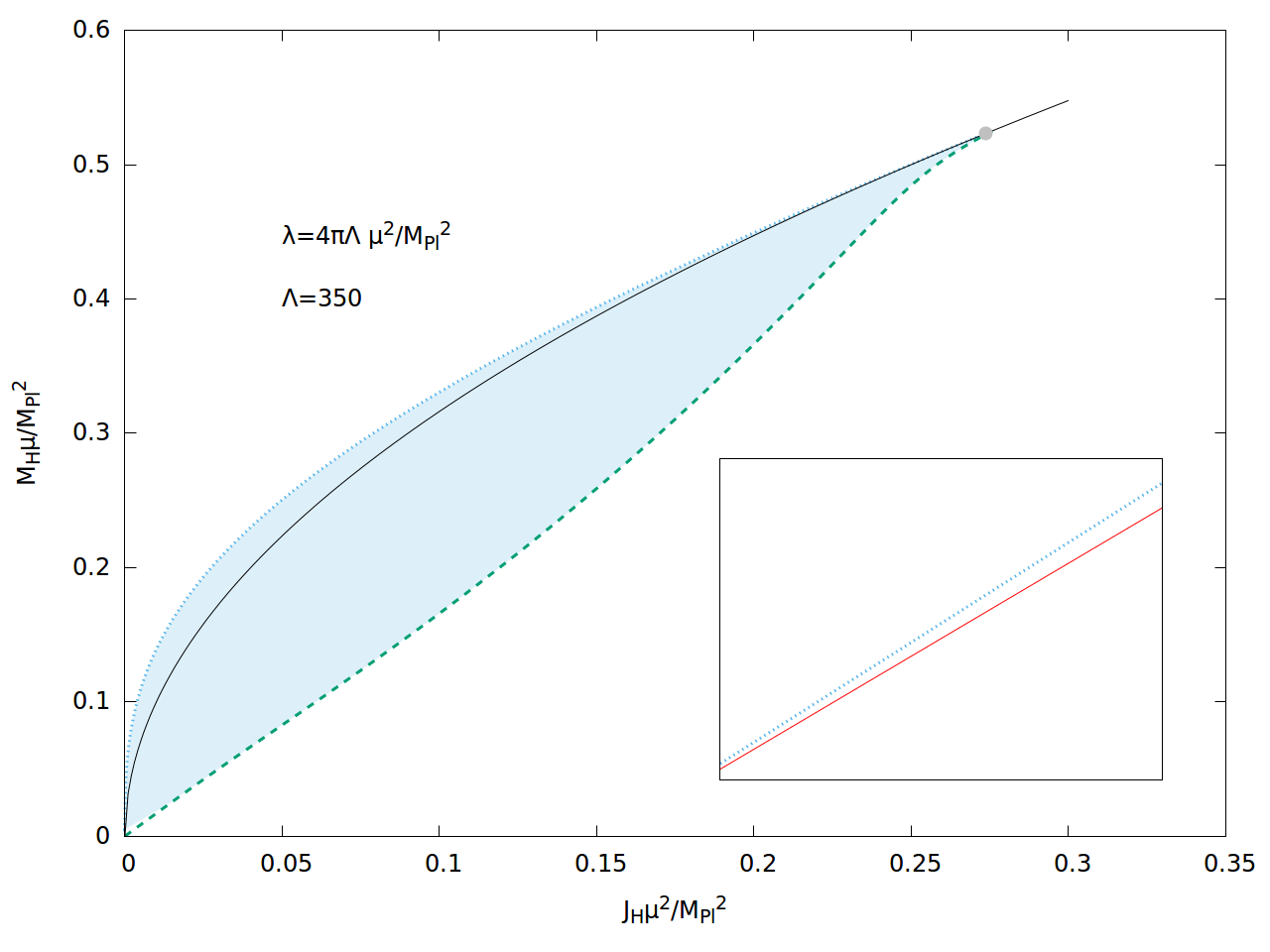}
      \end{center}
  \caption{Phase space in terms of horizon quantities.}
  \label{horizon_phase}
\end{figure}

We now turn our attention to the horizon mass $M_{\rm H}$ and angular momentum $J_{\rm H}$. In Fig.~\ref{horizon_phase}, we display the phase space of quartic-KBHsSH, for a particular value of $\lambda$, but that illustrates the general pattern we have observed.

In Fig.~\ref{horizon_phase}, we can see that all solutions lie within the envelope formed by the (blue dotted) line of vacuum Kerr BHs and the (green dashed) line of extremal KBHsSH, corresponding to the blue shaded region. The salient feature we wish to emphasize is that all quartic-KBHsSH have, therefore, a \textit{lower} mass than that at the Hod point. The same pattern is observed for all other values of $\lambda$ we have studied.

The black solid line in Fig.~\ref{horizon_phase} denotes the Kerr bound threshold, in terms of horizon quantities $J_{\rm H}=M_{\rm H}^2$ and it can be observed that there are solutions below it, thus violating the Kerr bound in terms of horizon quantities, as was observed in~\cite{Herdeiro:2015moa} for KBHsSH (without self-interactions). 

Finally, the inset of Fig.~\ref{horizon_phase} illustrates the typical pattern when starting from a given vacuum Kerr BH (on the blue dotted line) and increasing the hair, but keeping the frequency $w$ constant -- which we use as a control parameter. The corresponding points fall into the red solid line in the inset. One observes that making (vacuum) Kerr BHs hairier in this way, their horizon quantities increase, but the horizon mass increases more slowly in term of the horizon angular momentum than along the vacuum Kerr line.

\begin{figure}[h!]
  \begin{center}
    \includegraphics[width=8.55cm]{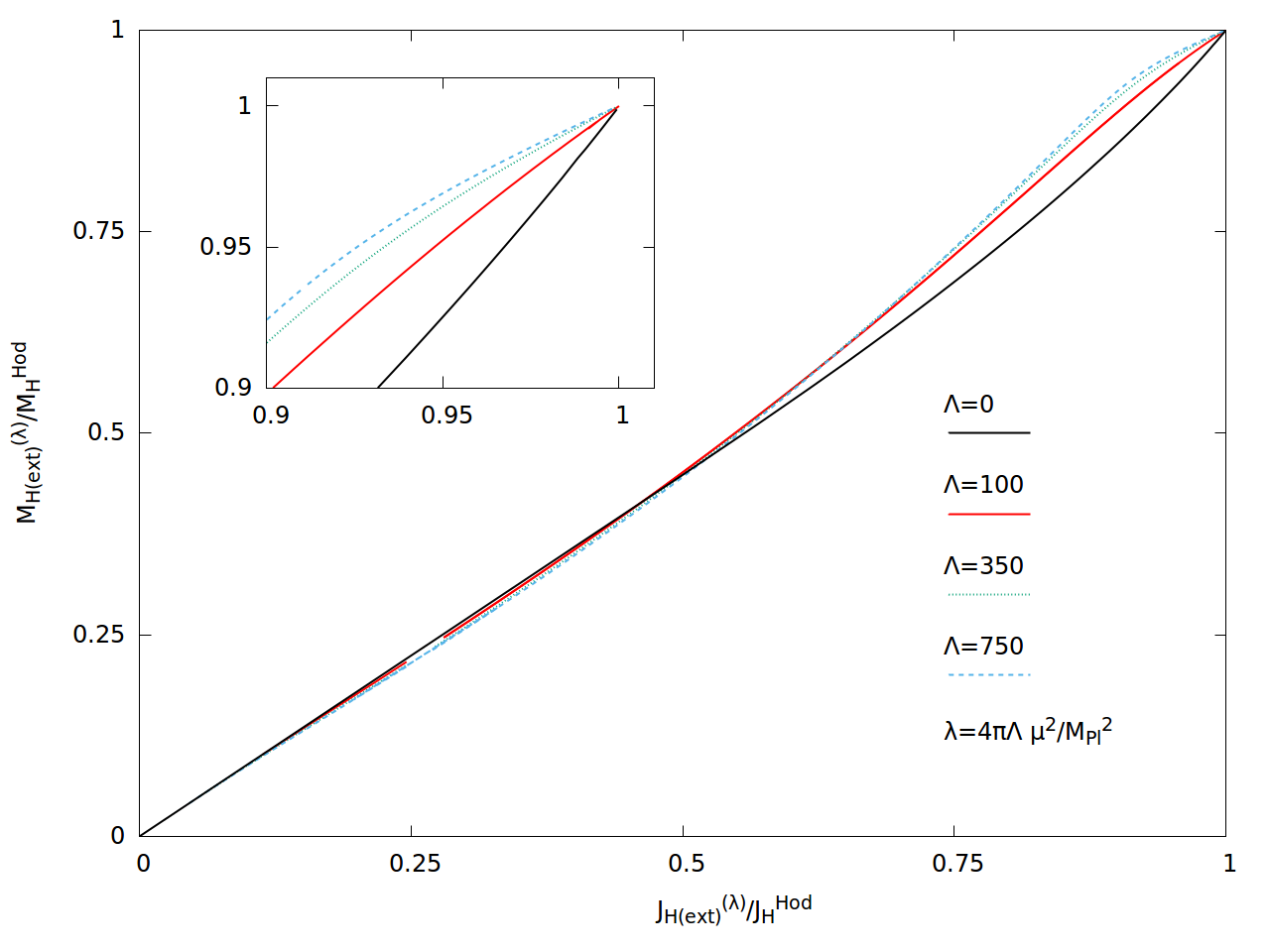}
      \end{center}
      \caption{ The horizon quantities of extremal  KBHsSH and quartic-KBHsSH given in terms of the values at the Hod point, for different values of $\lambda$.}
  \label{horizon_ratios}
\end{figure}

Fig.~\ref{horizon_phase} teaches us that the boundary of the domain of (quartic-)KBHsSH, in the phase space constructed from horizon quantities, is composed by a curve that does not depend on $\lambda$ and the extremal hairy BHs line, that depends on $\lambda$.  As such, we study in Fig.~\ref{horizon_ratios}
the ratio between $M_H^{(\lambda)}$, $J_H^{(\lambda)}$ for extremal BHs and the corresponding values at the Hod point. We see that this ratio, as a function of the ratio for the corresponding angular momenta, does not depend strongly on $\lambda$, and in particular is always a monotonic function. Moreover, as can be seen in the inset of this same figure, the maximal mass for extremal KBHsSH is obtained at the Hod point, as all the curves approach that point from below, regardless of the value of the coupling.

To summarize, as the (quartic-)KBHsSH are bounded by the extremal KBHsSH line and the vacuum Kerr BH line,  and since both of these curves have their maxima at the Hod point, 
we conclude that the maximal horizon quantities are obtained at the Hod point and all 
(quartic-)KBHsSH will have horizon quantities that are
lower than those at the Hod point.

\section{Discussion: universality and phenomenology}
\label{sec_V}
The main, and somehow unexpected, result in this paper is the observation that quartic-self interactions make KBHsSH ``hairier" but not ``heavier", in the sense of being able to increase their ADM mass but not their horizon mass, with respect to the maximal values attained in the absence of self-interactions. An immediate question is how \textit{universal} is this behaviour, if one considers a generic self-interacting scalar field theory.

To test this universality, we briefly consider a family of potentials allowing for localized solitonic configurations even in the absence of gravity. This family  considers hexic self-interactions and the corresponding scalar solitons -- known as \textit{Q-balls}~\cite{Coleman:1985ki,Lee:1991ax}, $i.e$ bundles of Noether charge associated with the $U(1)$ symmetry (typically denoted by $Q$)  -- have found a wide range of applications and interest in field theory (see $e.g.$~\cite{Kusenko:1997si,Enqvist:1997si}).  Thus, we replace the potential (\ref{pot}) by  
\begin{eqnarray}
U(|\Psi|)= \mu^2\left|\Psi\right|^2 - \lambda\left|\Psi\right|^4+  \beta\left|\Psi\right|^6,
\end{eqnarray}
where $\lambda>0$, $\beta>0$. In the following we take $\lambda=2$ and $\beta=1$ 
and it is convenient to introduce 
as well the dimensionless coupling constant $\alpha^2=4\pi G\mu/\sqrt{\beta}$.

On the one hand, a systematic study of the spinning solitons 
(including the effects induced by the backreaction on the spacetime geometry) 
is given in \cite{Kleihaus:2005me}.
The results therein show that the
 generic gravitating BSs in this theory, which we dub \textit{Q}-BSs, follow closely the pattern for mini-BSs and quartic-BSs described herein; in particular 
one finds a similar, albeit somewhat distorted, version of the mass $vs.$ frequency diagram shown in Fig.~\ref{fig:w-vs-M-BSs}.

\begin{figure}[h!]
  \begin{center}
     \includegraphics[width=8.55cm]{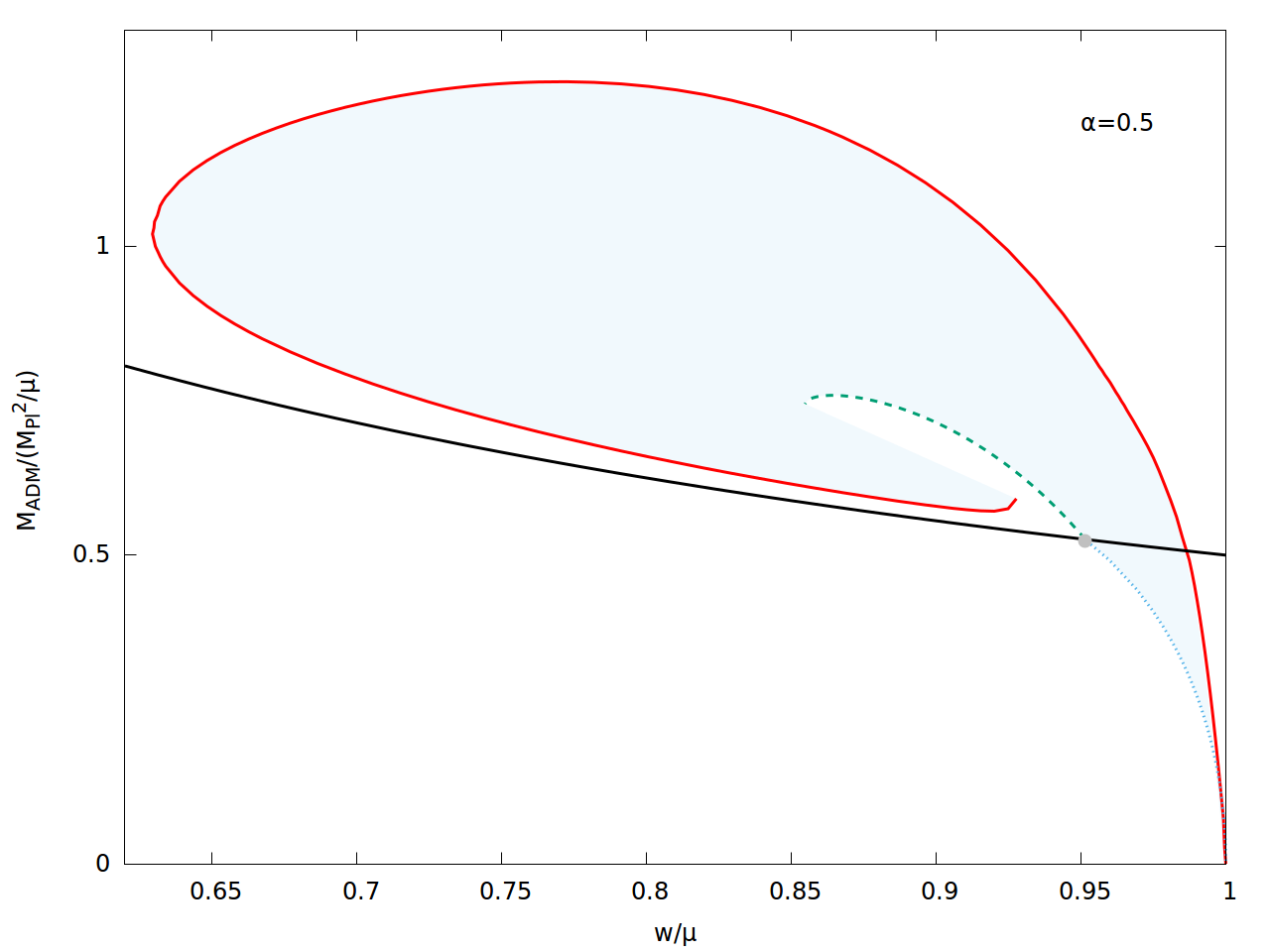}
      \end{center}
  \caption{Domain of existence of a typical set of $Q$-KBHsSH (blue shaded region).}
  \label{qkbhsh}
\end{figure}

On the other hand, Ref.~\cite{Herdeiro:2014pka} has studied $Q$-clouds, $i.e$ (non-backreacting) $Q$-balls on the (vacuum) Kerr BH background. We have done some preliminary studies of backreacting $Q$-balls, and found families of KBHsSH with this type of self-interactions, which we dub $Q$-KBHsSH. The overall properties of these BH solutions are similar to those found herein for  quartic-KBHSSH.
In particular, the domain of existence is still bounded by the (corresponding)
curves ${\bf i)}$, ${\bf ii)}$ and ${\bf iii)}$ as defined in Section \ref{sec_III}. 
Therefore, the maximal ADM mass 
of the solutions is 
that of the corresponding $Q$-BSs.
We remark that for the same scalar field mass, $Q$-KBHsSSH seem able to become considerably hairier than KBHsSH or quartic-KBHsSH.
An example of the domain of existence of $Q$-KBHsSH is presented in Fig.~\ref{qkbhsh} for a particular choice of couplings.

\begin{figure}[h!]
  \begin{center}
    \includegraphics[width=8.55cm]{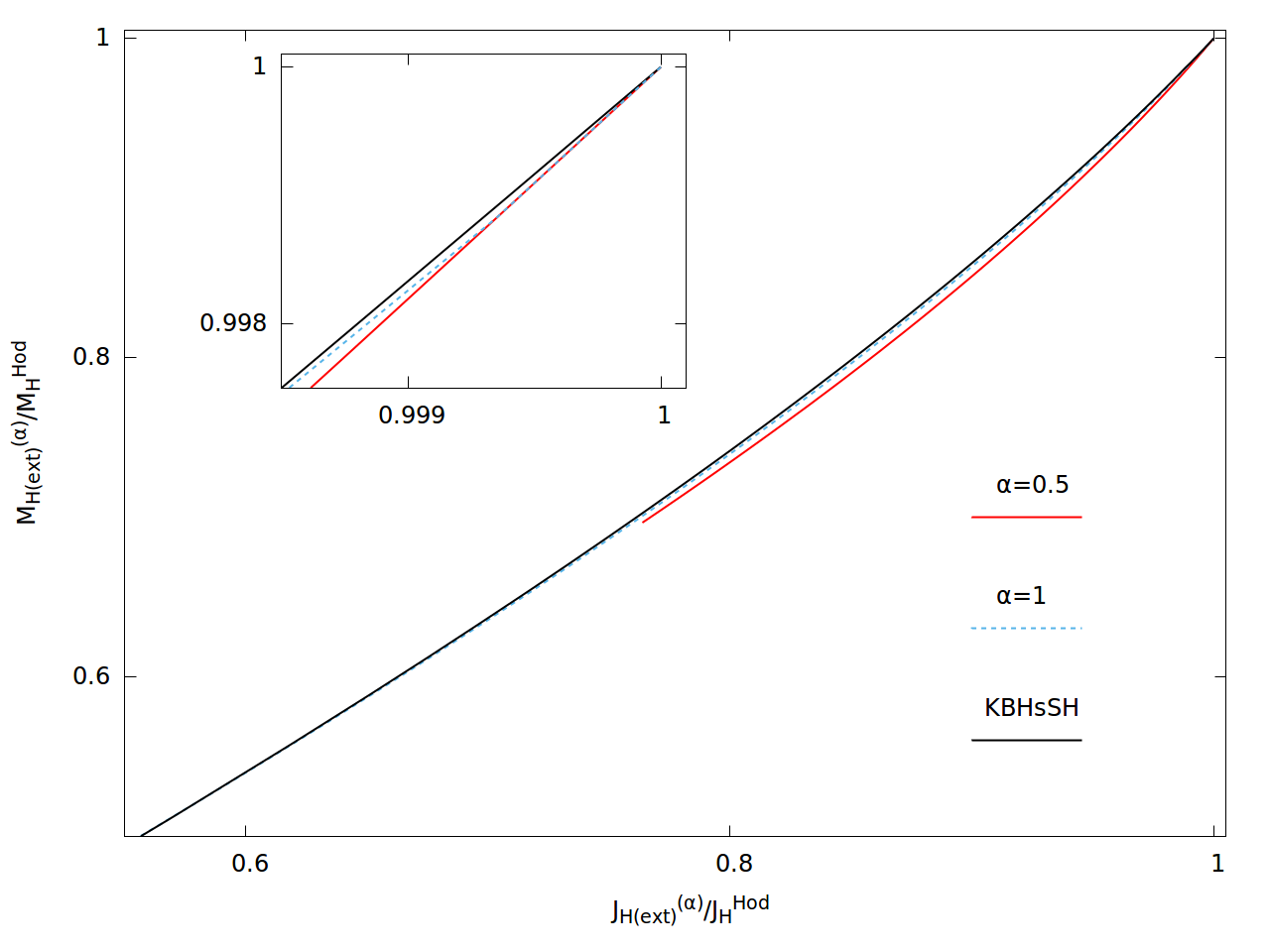}
      \end{center}
      \caption{ The horizon quantities of extremal  $Q$-KBHsSH given in terms of the values at the Hod point, for two different values of $\alpha$. The corresponding curve for KBHsSH without self-interactions almost overlaps with the $\alpha=1$ curve, confirming the behaviour described in the text.}
  \label{horizon_ratios_q}
\end{figure}

Concerning the horizon quantities, we have found
that for all families of solutions studied so far,
 the pattern in 
Figs. \ref{horizon_phase} and \ref{horizon_ratios},   
is preserved.
In particular, the maximum of the horizon mass and angular momentum
are those of the (universal) Hod point - Fig.~\ref{horizon_ratios_q}

A final observation concerning $Q$-KBHsSH, that follows from our numerical data 
is that 
when increasing the coupling
constant  $\alpha$,
the solutions tend to those with a free scalar field in \cite{Herdeiro:2014goa}. 
In fact, 
following the arguments in \cite{Kleihaus:2005me}, one can show that 
for large values of $\alpha$, all higher order terms in the scalar field potential become irrelevant
and, as $\alpha\to \infty$, one recovers the solutions of a model with a complex scalar field  
possessing a quadratic potential only. In fact, this feature starts to manifest already for $\alpha$  of order 1. This provides further evidence for the universality of the ``hairier" but not ``heavier" behaviour.

To close let us make a few remarks about phenomenology. To test the potential astrophysical relevance of KHBsSH, it would be interesting to compute, for both quartic and $Q$-KBHsSH, observables such as the frequency at the ISCO and the quadrupole moment. Both these properties have been discussed for KBHsSH without self-interactions in~\cite{Herdeiro:2014goa,Herdeiro:2015gia}, in terms of ADM quantities. The results in this paper suggest that large difference will occur when considering these properties in terms of ADM quantities, for the self-interacting case, whereas these observables should be mostly unchanged when considered in terms of horizon quantities, regardless of the self-interactions.  Another class of observables of interest concerns the BH shadows, in view of the ongoing and forthcoming observations of the Event Horizon Telescope. The shadows of KBHsSH (without self-interactions) have been recently studied~\cite{Cunha:2015yba} and can have strong deviations from the Kerr shadows. The strong deviations from the Kerr results are largely due to the lensing of the scalar ``hair" that surrounds the horizon. Thus, it seems reasonable to expect that such deviations can be aggravated for quartic or $Q$-KBHsSH which can become even ``hairier".

Finally, since the maximal horizon mass behaves as in \eqref{hodpoint}, if the ``hairier" but not ``heavier" property is universal and holds regardless of the scalar field theory, then KBHsSH with a trapped region with astrophysical masses can only exist if ultra-light scalars occur in Nature. This type of hairy BHs would be, therefore, a distinct observational signature of such beyond the Standard Model particles.

\vspace{0.5cm} 
\noindent
\section*{Acknowledgements}
C. H. and E. R. acknowledge funding from the FCT-IF programme. H.R. is supported by the grant PD/BD/109532/2015 under the MAP-Fis Ph.D. programme. This work was partially supported by the NRHEPÐ295189
FP7-PEOPLE-2011-IRSES Grant, by FCT via project No.
PTDC/FIS/116625/2010 and by the CIDMA strategic project UID/MAT/04106/2013. Computations were performed at the Blafis cluster, in Aveiro University.



\begin{small}

 \end{small}

\end{document}